\documentclass[twocolumn]{aastex62} 

\usepackage{graphicx}
\usepackage{natbib}

\newcommand{\igd}{$i_{\rm GD}$}
\newcommand{\iha}{$i_{\rm H\alpha}$}
\newcommand{\ha}{H$\alpha$}

\begin{document}

\title{Comparing Be Star Inclination Angles Determined from H$\alpha$ Fitting and Gravitational Darkening}

\accepted{for publication in the Astrophysical Journal, September 14, 2022.}

\correspondingauthor{T.\ A.\ A.\ Sigut}
\email{asigut@uwo.ca}

\author[0000-0002-0803-8615]{T.\ A.\ A.\ Sigut}
\affiliation{Department of Physics and Astronomy \\
The University of Western Ontario \\
London, Ontario, Canada N6A~3K7}
\affiliation{Institute for Earth and Space Exploration (IESX) \\
The University of Western Ontario}

\author[0000-0002-8693-3287]{Nastaran R. Ghafourian}
\affiliation{Department of Physics and Astronomy \\
The University of Western Ontario \\
London, Ontario, Canada N6A~3K7}

\begin{abstract}

Using a sample of 92 Galactic Be stars, we compare inclination angles (the angle between a star's rotation axis and the line-of-sight) determined from \ha\ emission line profile fitting to those determined by the spectroscopic signature of gravitational darkening. We find good agreement: 70\% of the sample (64 out of 92 stars) is consistent with zero difference between the two methods using $1\sigma$ errors, and there is a strong linear correlation coefficient between the two methods of $r=+0.63\pm0.05$. There is some evidence that the \ha\ profile fitting method overestimates the inclination angle for $i\lesssim 25^\circ$, perhaps due to the neglect of incoherent electron scattering on the \ha\ line widths, while the gravitational darkening method underestimates the inclination angle for $i\gtrsim70^\circ$, perhaps due to the neglect of disk radiative transfer effects on the optical spectrum. Overall, it is demonstrated that a single \ha\ spectrum of modest resolution and SNR can be used to extract a useful estimate for the inclination angle of an individual Be star. This allows equatorial rotation velocities for individual Be stars to be derived from $v \sin i $ measurements and will allow Be stars to be used to search for correlated spin axes in young, open clusters if unbiased (with respect to inclination) samples of Be stars are used.

\end{abstract}

\keywords{stars: rotation - (stars:) circumstellar matter - stars: emission-line, Be - stars: early-type, stars: fundamental parameters - - stars: formation - Galaxy: open clusters and associations}

\section{Introduction}

Open clusters in the Milky Way are important laboratories of high-mass star formation and new, large-scale photometric and spectroscopic surveys are revolutionizing our understanding of these objects \citep{Motte2018}. Massive star formation is a complex interplay of gravity, turbulence, rotation, radiation, and magnetic fields over many decades of length scale and gas density \citep{Tan2014,Rosen2020}. One particular facet is the flow of angular momentum from the parent molecular cloud down to the level of individual binary orbits and stellar spins \citep{Goodwin2007}. A possible way to constrain the relative importance of rotation (angular momentum) visa vie turbulence is to search for correlated spins among the stars formed in a cluster \citep{Corsaro2017,ReyRaposo2018}, and this can be done by measuring the distribution of stellar inclination angles. Here the inclination angle (usually denoted $i$) is the angle between a star's rotation axis and the line-of-sight. Isotropic stellar spins naturally produce a $p(i)\,di=\sin i\,di$ inclination distribution for the observer \citep{Gray2022}, and deviations from the $\sin i$ distribution can be used to detect and quantify correlated spins.

Several techniques have been used to determine inclination angles for individual stars: If the rotation period for the star can be unambiguously identified from photometric or spectroscopic time-series, this period can be combined with an estimate of the stellar radius and a measurement of the star's projected equatorial rotation speed (usually denoted $v\sin i$) to determine the inclination angle \citep{Abt1995,Kuszlewicz2019,Healy2020,Masuda2020}. Stellar rotation near the critical velocity (where the effective gravitational acceleration, $\log g$, at the star's equator vanishes) distorts the stellar surface and causes the photospheric parameters $(T_{\rm eff}, \log g)$ to vary with stellar latitude \citep{vonZeipel1924,Espinosa2011}; detailed spectroscopic modelling can then recover the viewing inclination \citep{Stoeckley1968,Fremat2005,Fremat2006,Zorec2016}. Asteroseismology is another powerful tool; if the rotational splitting of identified pulsation mode frequencies can be measured, the stellar inclination can be extracted \citep{Gizon2003,Corsaro2017,Mosser2018,Gehan2021}. Finally, matching the \ha\ emission line profile of a Be star to model template libraries can be used to determine the central B star's inclination angle \citep{Sigut2020}. This \ha\ template method is the focus of the current work and is discussed further below.

Using asteroseismology, \citet{Corsaro2017} derived inclination angles for 48 red giants in the old MW open clusters NCG~6791 (25 stars) and NGC~6819 (23 stars), finding evidence of strong spin correlation in both clusters, with stellar inclinations concentrated at the low inclination angles of $<\!\!i\!\!>\sim 30^\circ$ for NGC~6791 and $<\!\!i\!\!>\sim 20^\circ$ for NGC~6819. However, this result has been criticized by \citet{Mosser2018} and \citet{Gehan2021} who argue that the \citet{Corsaro2017} analysis favoured low inclination angles and high mode splittings.  A re-analysis by \citet{Mosser2018} and \citet{Gehan2021} find inclinations in both NGC~6791 and NGC~6819 consistent with the $\sin i$ distribution. Searches for correlated spins (using the rotation period method) in other MW clusters, such as NGC~2516 \citep{Healy2020} or M35 and the Pleiades and Praesepe open clusters \citep{Healy2021}, have found results consistent with both isotropic spins and moderate spin alignment.

Recently, \citet{Sigut2020} demonstrated that fitting a single-epoch, moderate resolution, moderate signal-to-noise \ha\ emission line profile of a Be star provides an efficient method to derive its inclination angle. On the main sequence, the B-type stars exhibit the highest rotation speeds, and rapid rotation is thought to be the driving factor in the ejection of a equatorial, circumstellar disk forming a B-emission line or Be star \citep{Struve1931,Porter2003,Rivinius2013}. The central B star's photospheric absorption spectrum is modified by the presence of the ionized, circumstellar disk in a Be star system and the morphology of the \ha\ emission line has long been known to reflect how the circumstellar disk is viewed \citep{Porter2003}. As the disk is equatorial, the axis of the disk immediately gives the axis of the central star. Be stars are quite common, representing $\approx\!20$\% of all main sequence B-type stars \citep{Zorec1997}, and \citet{Sigut2020} showed that moderate samples of 15 to 20 Be stars in a given cluster can be used to detect correlated spins in favourable cases.

\citet{Sigut2020} also compared inclination angles derived from \ha, \iha\  hereafter, with those derived from optical interferometry which directly resolves the major and minor axis of the disk emission on the sky. While the stellar sample was small, only 11 stars, agreement was excellent with a correlation coefficient of $r=+0.94$ between the two methods. It is the purpose of the present work to compare \iha\ to another, larger source of inclination angles for the Be stars, namely those based on gravitational darkening. 

\citet{Zorec2016} assembled a large sample of 233 classical Be stars in the Galactic field in order to study the distribution of their rotation speeds, and in particular, the distribution of rotation speeds relative to the critical speed,
\begin{equation}
\label{eq:vcrit}
v_{\rm crtic} = \left(\frac{GM}{1.5\,R}\right)^{1/2} = 357\,\left(\frac{M_*}{R_*}\right)^{1/2}\;{\rm km\,s^{-1}} \,,
\end{equation}
at which the effective gravitational acceleration at the stellar equator is zero \citep{Maeder2009book}. Here $M$ and $R$ are the stellar mass and radius and $M_*$ and $R_*$ are the stellar mass and radius in solar units. A by-product of this analysis was the determination of inclination angles for the sample stars. The method of \citet{Fremat2005,Fremat2006} was used to determine the {\it parent, nonrotating counterpart parameters\/} (pnrc) for each star. The derived pnrc parameters are the fundamental stellar parameters (mass, radius, luminosity) of a non-rotating star that when spun up to equatorial velocity $v$ and viewed at inclination angle $i$ best reproduces the observed spectrum and overall spectral energy distribution (SED). The main physical effect used in this analysis was gravitational darkening \citep{vonZeipel1924,Collins1965} caused by the Be star's presumed rapid rotation. Here, the local temperature variation with stellar latitude caused by rapid rotation makes the spectrum dependent on the viewing inclination. This idea, that the stellar inclination can be derived from the spectra of rapidly rotating, gravitationally darkened stars, goes back to \citet{Stoeckley1968}. For this reason, we will refer to the inclination angles derived by \citet{Zorec2016} as ``gravitational darkening" inclinations (or \igd) to distinguish them from the inclinations derived in the present work based on H$\alpha$ line profile fitting (\iha). It is important to note that many more ingredients beyond gravitational darkening went into the determination of the stellar inclination angles, and we refer the reader to \citet{Zorec2016} for details.

One additional point is in order: the inclination angles derived by \citet{Zorec2016} were a by-product of their analysis and were not used in their paper to derive equatorial velocities directly from measured $v\sin i$ values, likely because the inclination distribution was not consistent with the expected, random $\sin i$ distribution as there was a absence of stars with $i_{\rm GD}\gtrsim 70^\circ$. \citet{Zorec2016} instead chose to de-convolve the $v\sin i$ distribution assuming a random $\sin i$ distribution for the sample. Despite this, we feel that the \citet{Zorec2016} \igd\ inclination angles are a valuable resource to test the H$\alpha$ fitting method. We will also return to the question of a non-random distribution of inclination angles for this sample in a later section. 

\section{The Be Star Sample}

\begin{figure}
\begin{center}
\includegraphics[width=0.4\textwidth]{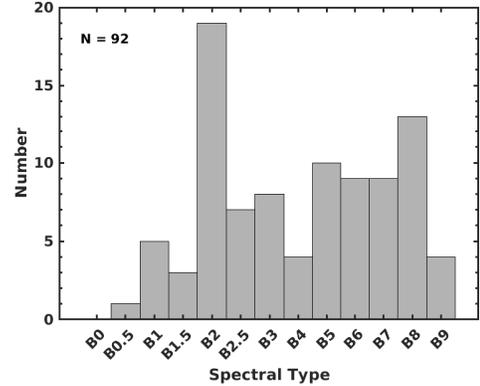}
\caption{Spectral type distribution of the Be star sample. \label{fig:spthist}}
\end{center}
\end{figure}

Of the 233 Be stars considered by \citet{Zorec2016}, we have obtained a sample of 92 stars with available \ha\ spectra, 58 from the BeSS spectral database\footnote{Operated at LESIA, Observatoire de Meudon, France: http://basebe.obspm.fr} and 34 from the John S.\ Hall telescope at Lowell Observatory \citep[see][]{Silaj2014}. The Hall spectra have ${\cal R}=10^4$ and SNR of $\approx 10^2$ or better, while the spectra from BeSS are a heterogeneous group with a range of SNRs and resolutions. The full sample is listed in Table~\ref{table:sample}; given for each star is the adopted spectra type from the SIMBAD database \citep{Wenger2000}, as well as the source (Hall or BeSS) and any particular notes about the spectrum. Figure~\ref{fig:spthist} shows a histogram of the sample spectral type. While spectral type B2 is the most populated bin, there is essentially an equal number of early-type (B3 or earlier) and late-type Be stars. Also given for each star in Table~\ref{table:sample} is the star's \ha\ shell parameter and its V/R ratio. The shell parameter is defined as the average of the violet (V) and red (R) emission peak fluxes in the \ha\ profile divided by the line-centre flux, and the V/R ratio is defined as the ratio of the peak V and R fluxes. The V/R ratio is a measure of the asymmetry in the line profile, while shell parameters exceeding 1.5 are taken to indicate shell stars with enhanced central absorption due to the line-of-sight passing through the disk \citep[see][for a careful discussion]{Hanuschik1996}. A singly-peaked profile is taken to have both a V/R ratio and shell parameter equal to one.

\begin{figure}
\begin{center}
\includegraphics[width=0.4\textwidth]{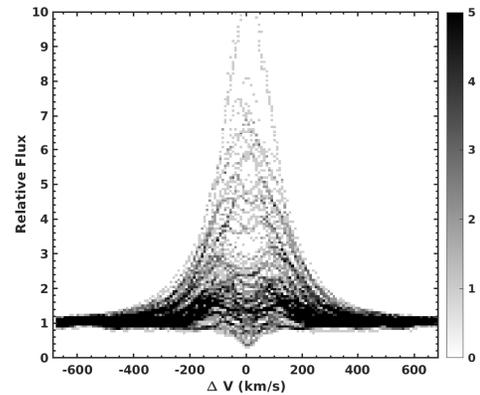}
\caption{Heat map of the sample \ha\ profiles. The colour bar indicates the number of line profiles passing through each box $(\Delta v,\Delta F)$, from a minimum of 0 to maximum of 92, the sample size (see text for details). The grey-scale is highly saturated, with entries from 5 through 92 assigned to black to bring out weaker features. \label{fig:heat}}
\end{center}
\end{figure}

Figure~\ref{fig:heat} shows a ``heat map" of the 92 sample \ha\ line profiles. Here the relative flux - wavelength (expressed as a velocity shift from line centre) plane is dividing into a 150-by-150 grid, and the number of line profiles passing through each box is counted; the counts (min 0, max 92) in each box are then displayed as grey-scale. A wide range of profile shapes are represented, from single emission peaks, to doubly-peaked profiles, to doubly-peaked profiles with deep shell absorption. There is no claim that this is an unbiased sample of Be stars; the only requirement of the present work is that a wide range of inclination angles be present in the sample, something demonstrated by the wide range of profile morphologies in Figure~\ref{fig:heat}.

\begin{figure}
\begin{center}
\includegraphics[width=0.4\textwidth]{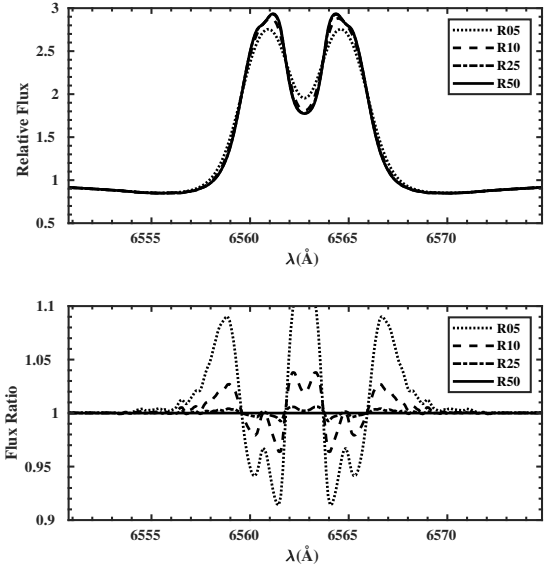}
\caption{{\it Top panel:} effect of spectral resolution on the \ha\ relative flux profile for the disk density model $(n,\rho_0,R_d,i)=[3.0,7.69\times 10^{-11}{\rm g\,cm^{-3}},25R_*,60^\circ]$. Shown are spectral resolutions $\mathcal R=5000,$ (dotted) $10,000$ (dashed), $25,000$ (dash-dotted) and $50,000$ (solid). {\it Bottom panel:} relative flux profile ratios, with each profile of the top panel divided by the line profile for $\mathcal R=50,000$. \label{fig:reseg}}
\end{center}
\end{figure}

One issue with the BeSS spectra is their heterogeneity. Visually, all of the spectra have SNRs exceeding 50 and resolutions of at least 10,000. Ideally it would be best to match the spectral resolution of each observed profile with the \ha\ profile library used to extract the inclination. Unfortunately, the resolution given in many of the \texttt{fits} headers seemed incorrect or implausible. Visually, a few profiles appeared to have resolutions of at least 50,000. In practice, Be star \ha\ profiles change little beyond a resolution of 10,000 or so. This is shown in Figure~\ref{fig:reseg} which compares theoretical \ha\ line profiles computed at resolutions from  5,000 to 50,000. As can be seen from the figure, after a resolution of 10,000, there is little change in the profile ($\lesssim 5$ percent) and after 25,000, the changes are negligible. Both resolution 10,000 and 25,000 \ha\ profiles will be used to fit the BeSS sample stars to gauge the inclination errors introduced by the uncertainty in resolution.

All of the sample \ha\ line profiles were visually shifted to their rest frame corresponding to a line-centre wavelength of $\lambda\,6562.8\,$\AA. The line fitting procedure (described in the next section) also included small additional shifts of $\pm 0.5\,$\AA\ to optimize the fits. For each BeSS spectrum, the continuum normalization was checked and adjusted, if required (see Table~\ref{table:sample}), using a straight-line fit between the adjacent blue and red continua. In all cases, these adjustments were small. The shell parameters and V/R ratios of Table~\ref{table:sample} were all measured from the spectra after any adjustment.  

\startlongtable
\begin{deluxetable*}{rllrrrrrll}
\tablecaption{The Be star sample.\label{table:sample}}
\tablecolumns{10}
\tablehead{
\colhead{HD} &  \colhead{Star} & \colhead{Spectral} & \colhead{$i_{\rm H\alpha}\pm 1\sigma$} & \colhead{$i_{\rm GD}\pm 1\sigma$}\tablenotemark{a} & \colhead{Shell} &  \colhead{V/R} & \colhead{$v\sin i \pm 1\sigma$}\tablenotemark{a} & \colhead{Source} & \colhead{Notes} \\
\colhead{~}  &  \colhead{~} & \colhead{Type} & \colhead{$(deg)$} & \colhead{$(deg)$} & \colhead{Param} & \colhead{~} & \colhead{($\rm km\,s^{-1}$)} & \colhead{~} & \colhead{~} }
\startdata
     144 & * 10 Cas     & B9III   & $21\pm  5$ & $38\pm  9$ &  1.04 &  1.01 & $140\pm 11$ & Hall  & \\        
    4180 & * omi Cas    & B5III   & $60\pm 11$ & $58\pm 14$ &  1.00 &  1.00 & $215\pm 18$ & Hall  & \\        
    5394 & gam Cas      & B0.5IV  & $59\pm 12$ & $65\pm 16$ &  1.00 &  1.01 & $441\pm 51$ & Hall  & \\        
    6811 & * phi And    & B5III   & $12\pm  5$ & $29\pm  7$ &  1.00 &  1.00 & $ 95\pm 13$ & Hall  & \\        
   10144 & alf Eri      & B6V     & $70\pm  7$ & $69\pm 17$ &  1.38 &  1.02 & $255\pm 35$ & BeSS  & T\\        
   10516 & * phi Per    & B1.5V   & $69\pm  6$ & $57\pm 14$ &  1.11 &  1.10 & $603\pm 55$ & Hall  & \\        
   18552 & HD 18552     & B7IV    & $78\pm 14$ & $72\pm 18$ &  1.26 &  1.00 & $294\pm 37$ & BeSS  & \\        
   20336 & HD 20336     & B2.5V   & $63\pm  6$ & $68\pm 17$ &  1.13 &  1.01 & $346\pm 39$ & Hall  & \\        
   22192 & psi Per      & B5V     & $74\pm  5$ & $74\pm 18$ &  2.17 &  1.01 & $306\pm 39$ & Hall  & \\        
   22780 & HD 22780     & B7V     & $73\pm  5$ & $89\pm 22$ &  1.31 &  1.01 & $306\pm 35$ & BeSS  & \\        
   23016 & 13 Tau       & B7V     & $28\pm  5$ & $59\pm 14$ &  1.18 &  1.01 & $273\pm 30$ & BeSS  & \\        
   23302 & 17 Tau       & B6III   & $20\pm  5$ & $45\pm 11$ &  1.12 &  1.01 & $198\pm 26$ & BeSS  & T,N\\        
   23480 & 23 Tau       & B6IV    & $37\pm  5$ & $70\pm 17$ &  1.21 &  1.00 & $257\pm 36$ & Hall  & \\        
   23552 & HD 23552     & B8V     & $48\pm  9$ & $65\pm 16$ &  1.09 &  1.03 & $236\pm 29$ & Hall  & \\        
   23630 & eta Tau      & B7III   & $43\pm  5$ & $62\pm 15$ &  1.03 &  1.01 & $150\pm 21$ & Hall  & \\        
   23862 & 28 Tau       & B8V     & $78\pm  5$ & $68\pm 17$ &  2.45 &  1.05 & $332\pm 33$ & Hall  & \\        
   25940 & *c Per       & B3V     & $50\pm 14$ & $50\pm 12$ &  1.00 &  1.00 & $219\pm 29$ & Hall  & \\        
   28497 & * 228 Eri    & B2V     & $68\pm  6$ & $56\pm 14$ &  1.25 &  1.09 & $362\pm 43$ & BeSS  & \\        
   32343 & * 11 Cam     & B3V     & $37\pm  9$ & $24\pm  8$ &  1.00 &  1.00 & $105\pm 10$ & Hall  & \\        
   33453 & HD 33453     & B8V     & $55\pm  5$ & $89\pm 22$ &  1.26 &  1.00 & $290\pm 39$ & BeSS  & N\\        
   35439 & * psi01 Ori  & B1V     & $63\pm  5$ & $58\pm 14$ &  1.17 &  1.03 & $284\pm 28$ & Hall  & \\        
   36012 & HD 36012     & B2V     & $41\pm  5$ & $41\pm 12$ &  1.00 &  1.00 & $195\pm 22$ & BeSS  & \\        
   36576 & * 120 Tau    & B2IV    & $61\pm  9$ & $57\pm 14$ &  1.09 &  1.03 & $282\pm 29$ & BeSS  & 25\\      
   37490 & * ome Ori    & B3V     & $59\pm 10$ & $50\pm 12$ &  1.15 &  1.02 & $190\pm 21$ & BeSS  & 25\\      
   37795 & alf Col      & B9V     & $43\pm  6$ & $55\pm 13$ &  1.13 &  1.00 & $199\pm 17$ & BeSS  & \\        
   37967 & HD 37967     & B2.5V   & $56\pm 12$ & $52\pm 13$ &  1.00 &  1.00 & $234\pm 27$ & BeSS  & \\        
   41335 & HD 41335     & B1V     & $78\pm  6$ & $88\pm 22$ &  1.25 &  1.04 & $384\pm 50$ & BeSS  & \\        
   42545 & f01 Ori      & B5V     & $50\pm  5$ & $64\pm 16$ &  1.17 &  1.02 & $312\pm 34$ & BeSS  & \\        
   43544 & HD 43544     & B2V     & $55\pm  6$ & $46\pm 11$ &  1.32 &  1.01 & $274\pm 25$ & BeSS  & N\\        
   44996 & HD 44996     & B2.5V   & $46\pm  5$ & $13\pm  9$ &  1.17 &  1.03 & $ 40\pm  3$ & BeSS  & \\        
   45542 & nu Gem       & B6IV    & $68\pm  8$ & $69\pm 17$ &  2.14 &  1.08 & $233\pm 27$ & Hall  & \\        
   45871 & HD 45871     & B3V     & $78\pm  5$ & $56\pm 14$ &  1.77 &  1.03 & $299\pm 37$ & BeSS  & T,N\\        
   47054 & HD 47054     & B8IV    & $51\pm  5$ & $67\pm 16$ &  1.16 &  1.00 & $231\pm 32$ & BeSS  & \\        
   50013 & * kap Cma    & B1.5V   & $57\pm 13$ & $50\pm 17$ &  1.14 &  1.20 & $231\pm 25$ & Hall  & \\        
   56139 & * ome Cma    & B2.5V   & $37\pm  5$ & $24\pm  8$ &  1.00 &  1.00 & $ 94\pm 13$ & Hall  & \\        
   58050 & HD 58050     & B2V     & $21\pm  6$ & $31\pm 12$ &  1.31 &  1.02 & $138\pm 12$ & BeSS  & N\\        
   58343 & HD 58343     & B2V     & $29\pm  8$ & $17\pm 10$ &  1.00 &  1.00 & $ 47\pm  6$ & Hall  & \\        
   58715 & bet Cmi      & B8V     & $49\pm  6$ & $66\pm 16$ &  1.12 &  1.02 & $239\pm 28$ & Hall  & \\        
   60606 & * z Pup      & B2V     & $55\pm  5$ & $63\pm 15$ &  1.47 &  1.06 & $291\pm 34$ & BeSS  & 25\\      
   68980 & * r Pup      & B1V     & $43\pm  7$ & $33\pm 19$ &  1.00 &  1.00 & $149\pm 20$ & Hall  & \\        
   77320 & HD 77320     & B2V     & $63\pm  5$ & $65\pm 16$ &  1.27 &  1.01 & $364\pm 35$ & BeSS  & \\        
   83953 & *I Hya       & B5V     & $78\pm  7$ & $64\pm 16$ &  1.12 &  1.03 & $285\pm 30$ & Hall  & \\        
   86612 & HD 86612     & B5V     & $60\pm 10$ & $36\pm  9$ &  1.00 &  1.00 & $198\pm 17$ & Hall  & \\        
   88661 & HD 88661     & B5V     & $66\pm  6$ & $46\pm 11$ &  1.00 &  1.00 & $250\pm 30$ & BeSS  & \\        
   89080 & ome Car      & B7III   & $59\pm  7$ & $77\pm 19$ &  1.52 &  1.01 & $257\pm 30$ & BeSS  & T,25\\      
   91120 & HD 91120     & B8IV    & $56\pm  7$ & $89\pm 22$ &  1.29 &  1.00 & $329\pm 29$ & Hall  & \\        
   91465 & p Car        & B4V     & $64\pm  6$ & $73\pm 18$ &  1.09 &  1.12 & $286\pm 24$ & BeSS  & \\        
  105435 & del Cen      & B2V     & $57\pm 12$ & $52\pm 13$ &  1.06 &  1.04 & $263\pm 30$ & BeSS  & \\        
  105521 & HD 105521    & B3IV    & $24\pm  5$ & $42\pm 11$ &  1.20 &  1.01 & $152\pm 19$ & BeSS  & T\\        
  112091 & mu.02 Cru    & B5V     & $50\pm  7$ & $48\pm 15$ &  1.06 &  1.01 & $224\pm 20$ & BeSS  & \\        
  120324 & mu Cen       & B2V     & $38\pm 11$ & $30\pm 11$ &  1.06 &  1.06 & $149\pm 13$ & Hall  & \\        
  124367 & HD 124367    & B4V     & $64\pm  5$ & $63\pm 15$ &  1.07 &  1.03 & $321\pm 40$ & BeSS  & \\        
  127972 & eta Cen      & B2V     & $77\pm  5$ & $67\pm 16$ &  1.69 &  1.09 & $328\pm 33$ & Hall  & \\        
  131492 & * tet Cir    & B2III   & $60\pm  5$ & $39\pm  9$ &  1.12 &  1.01 & $196\pm 17$ & BeSS  & N\\        
  134481 & kap Lup      & B9V     & $21\pm  5$ & $45\pm 11$ &  1.09 &  1.04 & $177\pm 24$ & BeSS  & N\\        
  135734 & mu Lup       & B8V     & $58\pm  8$ & $70\pm 17$ &  1.25 &  1.01 & $285\pm 36$ & BeSS  & \\        
  137387 & * kap01 Aps  & B2V     & $84\pm  5$ & $50\pm 12$ &  2.42 &  1.01 & $264\pm 30$ & BeSS  & N\\        
  142926 & 4 Her        & B9V     & $77\pm  5$ & $89\pm 22$ &  1.70 &  1.03 & $360\pm 42$ & Hall  & \\        
  148184 & chi Oph      & B2V     & $51\pm 17$ & $23\pm 11$ &  1.00 &  1.00 & $149\pm 21$ & Hall  & \\        
  156325 & HD 156325    & B6V     & $54\pm  5$ & $53\pm 13$ &  1.34 &  1.03 & $199\pm 23$ & BeSS  & T\\        
  157042 & * iot Ara    & B1V     & $69\pm  5$ & $59\pm 14$ &  1.22 &  1.09 & $355\pm 46$ & BeSS  & \\        
  158427 & alf Ara      & B2V     & $66\pm  5$ & $56\pm 14$ &  1.09 &  1.00 & $299\pm 28$ & BeSS  & \\        
  162732 & * z Her      & B6III   & $86\pm  5$ & $62\pm 15$ &  5.40 &  1.06 & $353\pm 40$ & BeSS  & \\        
  164284 & 66 Oph       & B2V     & $35\pm  5$ & $51\pm 12$ &  1.17 &  1.01 & $264\pm 24$ & Hall  & \\        
  167128 & HD 167128    & B2V     & $44\pm 17$ & $15\pm  8$ &  1.17 &  1.05 & $ 53\pm  7$ & BeSS  & T\\        
  171219 & HD 171219    & B8V     & $78\pm  5$ & $69\pm 17$ &  6.49 &  1.09 & $199\pm 20$ & BeSS  & N\\        
  175869 & * 64 Ser     & B8II    & $39\pm  5$ & $60\pm 15$ &  1.08 &  1.02 & $184\pm 21$ & BeSS  & \\        
  183914 & bet Cyg B    & B8V     & $47\pm  9$ & $58\pm 14$ &  1.11 &  1.01 & $239\pm 31$ & BeSS  & \\        
  185037 & 11 Cyg       & B8V     & $65\pm  6$ & $70\pm 17$ &  1.22 &  1.00 & $286\pm 28$ & BeSS  & T,N,25\\        
  187811 & 12 Vul       & B2.5V   & $60\pm  5$ & $52\pm 13$ &  1.15 &  1.02 & $264\pm 25$ & BeSS  & \\        
  189687 & * 25 Cyg     & B3IV    & $53\pm 11$ & $49\pm 12$ &  1.13 &  1.05 & $215\pm 28$ & BeSS  & \\        
  191610 & b02 Cyg      & B2.5V   & $40\pm  5$ & $69\pm 17$ &  1.21 &  1.01 & $314\pm 33$ & Hall  & \\        
  192044 & * 20 Vul     & B7V     & $59\pm  8$ & $66\pm 16$ &  1.13 &  1.02 & $266\pm 35$ & BeSS  & \\        
  193911 & * 25 Vul     & B6IV    & $45\pm  8$ & $53\pm 13$ &  1.04 &  1.01 & $177\pm 21$ & BeSS  & \\        
  194335 & HD 194335    & B2III   & $54\pm  5$ & $60\pm 15$ &  1.24 &  1.00 & $374\pm 52$ & BeSS  & N\\        
  198183 & lam Cyg      & B5V     & $16\pm  5$ & $35\pm  8$ &  1.09 &  1.01 & $138\pm 17$ & BeSS  & N,T\\        
  200120 & * f01 Cyg    & B1.5V   & $67\pm  7$ & $89\pm 22$ &  1.06 &  1.02 & $407\pm 37$ & BeSS  & \\        
  201733 & HD 201733    & B4IV    & $75\pm  5$ & $72\pm 18$ &  1.54 &  1.01 & $363\pm 49$ & BeSS  & N\\        
  202904 & ups Cyg      & B2V     & $50\pm 13$ & $38\pm  9$ &  1.00 &  1.00 & $193\pm 23$ & Hall  & \\        
  205637 & * eps Cap    & B3V     & $83\pm  5$ & $68\pm 17$ &  2.43 &  1.01 & $258\pm 34$ & BeSS  & \\        
  208886 & HD 208886    & B7V     & $84\pm  5$ & $54\pm 13$ & 12.02 &  1.02 & $221\pm 20$ & BeSS  & \\        
  209014 & * eta PsA    & B8III   & $83\pm  8$ & $89\pm 22$ &  1.30 &  1.02 & $320\pm 32$ & BeSS  & \\        
  209409 & omi Aqr      & B5V     & $75\pm  5$ & $73\pm 18$ &  2.33 &  1.00 & $289\pm 33$ & Hall  & \\        
  210129 & * 25 Peg     & B7V     & $38\pm  6$ & $30\pm  7$ &  1.03 &  1.00 & $141\pm 14$ & Hall  & \\        
  212076 & * 31 Peg     & B3III   & $35\pm  8$ & $26\pm  9$ &  1.03 &  1.01 & $105\pm 13$ & BeSS  & \\        
  212571 & * pi. Aqr    & B1IV    & $56\pm  5$ & $43\pm 10$ &  1.14 &  1.01 & $242\pm 25$ & Hall  & \\        
  214748 & eps PsA      & B7III   & $42\pm  5$ & $60\pm 15$ &  1.12 &  1.01 & $213\pm 23$ & BeSS  & \\        
  217543 & HD 217543    & B2.5V   & $75\pm  5$ & $82\pm 20$ &  1.78 &  1.02 & $354\pm 32$ & BeSS  & T\\        
  217675 & * omi And    & B6V     & $88\pm  5$ & $89\pm 22$ &  3.50 &  1.06 & $301\pm 34$ & BeSS  & T\\        
  217891 & bet Psc      & B6V     & $31\pm  6$ & $26\pm 10$ &  1.00 &  1.00 & $ 98\pm  9$ & Hall  & \\        
  224559 & HD 224559    & B4V     & $63\pm  8$ & $72\pm 18$ &  1.16 &  1.01 & $316\pm 37$ & BeSS  & T,N\\        
  224686 & eps Tuc      & B8V     & $74\pm  5$ & $89\pm 22$ &  1.51 &  1.01 & $322\pm 42$ & BeSS  & T,N\\ \hline 
\enddata
\tablenotetext{a}{From Table~4 of \protect\cite{Zorec2016}.}
\tablecomments{The shell parameter (column 6) and V/R ratio (column 7) were measured in this work. In the Notes columns for each star: 25 indicates the ${\cal R}=25,000$ resolution library profiles were used; N indicates the observed profile was noisy; T indicates significant telluric features in the observed profile. }
\end{deluxetable*}

\section{Methods}

To extract the stellar inclination for each star of Table~\ref{table:sample} from \ha\ line profile fitting, we follow the procedure of \cite{Sigut2020} using the \texttt{Bedisk} and \texttt{Beray} circumstellar disk codes \citep{Sigut2018}. A mass, radius, and luminosity was assumed for each star based on its spectral type \citep[see Table~1 of][]{Sigut2020} and a large library of line profiles for this mass and radius was compared to the observed profile to find the best-fitting profile (see below). For a given mass, radius, and luminosity, each library \ha\ line profile is specified by four parameters, three $(n,\rho_0,R_d)$ fixing the equatorial disk density distribution in the expression
\begin{equation}
\label{eq:rho}
\rho(R,Z) \equiv \rho_0\,\left(\frac{R_*}{R}\right)^n\;\exp\left(-\left[\frac{Z}{H(R)}\right]^2\right)\;.
\end{equation}
and one fixing the viewing inclination, the angle $i$, which is the target of the fitting procedure. In Eq.~(\ref{eq:rho}), $(R,Z)$ are the cylindrical co-ordinates of the disk and the density is truncated at the assumed disk radius $R_d$. The scale height $H(R)$ satisfies
\begin{equation}
\label{eq:H}
\frac{H(R)}{R}=\frac{c_s(0.6\,T_{\rm eff})}{V_K(R)}\,
\end{equation}
where $c_s$ is the sound speed in the disk, computed for a temperature of 60\% of the central star's $T_{\rm eff}$, and $V_K$ is the Keplerian orbital speed at $R$, {\it i.e.\/} $\sqrt{GM_*/R}$ where $M_*$ is the adopted the stellar mass. Despite computing the sound speed used in the scale height at a fixed temperature representative of the disk, the full disk temperature structure $T(R,Z)$ is found by \texttt{Bedisk} by enforcing radiative equilibrium at a grid of locations within the disk, and this temperature structure is used in the radiative transfer solutions performed by \texttt{Beray}. For a discussion of disk models in consistent, vertical hydrostatic equilibrium where the sound speed is computed using the disk's temperature structure, see \citet{Sigut2009}; the effect of this refinement to the disk's vertical density structure on the predicted \ha\ profiles are small (however, see the discussion in Section~7).

To compute the \ha\ profiles, the \texttt{Beray} code was used which
requires as input the temperature $T(R,Z)$ and density $\rho(R,Z)$
structure of the disk as computed by \texttt{Bedisk}. \texttt{Beray}
solves the transfer equation along a large number of rays directed at the
observer. Rays terminating on the stellar surface use a Doppler-shifted,
photospheric \ha\ profile as the ``upwind" boundary condition, while rays
that pass through the disk and do not intersect the stellar surface use
zero intensity as the upwind boundary condition. With this procedure, a
consistent \ha\ profile is obtained for the combined star+disk system: in
the limit $\rho_0\rightarrow 0$, \texttt{Beray} will return the
photospheric profile appropriate to the star's spectral type, while
for $\rho_0\ne 0$, the effects of disk emission and absorption are
consistently computed.

The best-fit \ha\ profile was defined by minimizing the figure-of-merit
\begin{equation}
\label{eq:fom}
{\mathcal F} \equiv \frac{1}{W}\,\sum_{i=1}^{i=N} w_i\,\frac{|F_i^{\rm mod} - F_i^{\rm obs}|}{F_i^{\rm obs}}\;,
\end{equation}
where $F_i^{\rm mod}$ and $F_i^{\rm obs}$ are the model and observed continuum-normalized fluxes at wavelength $\lambda_i$ respectively, $w_i$ is a weight (discussed below), and $W$ is equal to the sum of the weights $w_i$. The sum in Eq.~(\ref{eq:fom}) is over the $N$ wavelengths within the interval to be fit. Ranges of $\pm 12$ and $\pm 15\,$\AA\ from line centre (or $\pm548$ and $\pm686\,\rm km\,s^{-1}$ respectively) were considered to gauge the effect on the results. Following \cite{sigut2015,Sigut2020}, two choices were considered for the weights $w_i$, uniform-weighting with $w_i=1$ (and $W=N$) and core-weighting with $w_i=|F_i^{\rm obs}-1|$. The latter clearly weights the emission peaks more than fluxes near the continuum. These two possibilities for the figure-of-merit weighting plus the two wavelength ranges for the fitting region made four ``hyper-parameter" combinations for the H$\alpha$ fitting method. More details on the construction of the \ha\ line profile grids and the range of disk parameters $(\rho_0,n,R_d)$ considered can be found in \citet{Sigut2020}.

All four hyper-parameter combinations were used to fit each observed \ha\ profile. Each hyper-parameter choice produces a single, best-fit model to the observed profiles with Eq.~(\ref{eq:fom}) giving ${\mathcal F}={\mathcal F}_{\rm min}$ and the corresponding set of best-fit model parameters $(\rho_0,n,R_d,i)_{\rm best}$. However, there are also profiles in the library, corresponding to slightly different model parameters $(\rho_0,n,R_d,i)$, that fit the observed profile nearly as well as the best-fit profile. We define this set as the parameters of those model profiles with fitting figure-of-merits ${\cal F}$ that satisfy ${\mathcal F}\le 1.15 {\mathcal F}_{\rm min}$; the uncertainty in each model parameter is then the standard deviation of that parameter over this set of models. Focusing on just the inclination, this gives an estimate $i\pm 1\sigma$ for each of the four hyper-parameter fits. The average of these four inclination is taken to be the \iha\ inclination, listed in Table~\ref{table:sample}. The average of the four error estimates, $\bar{\sigma}$, is compared to the standard deviation of the best inclinations found from the four hyper-parameter fits, $\sigma_4$. If $\bar{\sigma}>\sigma_4$, $\bar{\sigma}$ is adopted as the error, otherwise $\sigma_4$ is adopted as the error. In all cases, a minimum error of $\pm 5^\circ$ was enforced. This process defines the \iha\ estimate for each stars and its error. It is a feature of Be star \ha\ line profiles that similarly fitting models usually all share very similar inclinations \citep[see][]{Sigut2020}; if this is not the case for a particular observed profile, the described procedure to assign the inclination error will assign a large uncertainty to \iha. 
Finally, estimates of the other model parameters $(\rho_0,n,R_d)$ are also generated for each star, but they are not the focus of this work. The distribution of these parameters for the 92 star sample is briefly discussed in Section~\ref{sec:otherparam}.

\begin{figure}
\begin{center}
\includegraphics[width=0.5\textwidth]{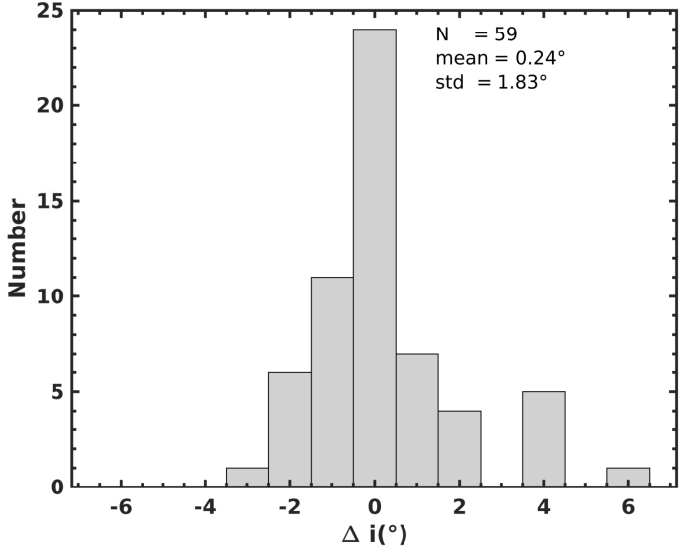}
\caption{Histogram of \ha\ inclination angle differences for the 59 stars with BeSS spectra when $\mathcal R=10,000$ versus $\mathcal R=25,000$ library profiles are used for the fit. $\Delta i \equiv i_{10} - i_{25}$ and is in degrees. \label{fig:r10v25}}
\end{center}
\end{figure}

The entire sample was first fit with $\mathcal R=10,000$ library profiles. As noted, however, the exact resolution of many of the BeSS spectra was not clear from the \texttt{fits} headers. To address this, the 59 stars with BeSS spectra were refit with $\mathcal R=25,000$ library profiles. A histogram of the differences between $\mathcal R=10,000$ and $R=25,000$ derived inclination angles is shown in Figure~\ref{fig:r10v25}. The inclinations derived from the two resolutions agree to within $\pm2^\circ$ for 52 out of the 59 stars; six of the remaining stars agree to within $\pm4^\circ$, and a single star, HD~89080, differs by $6^\circ$. Visual inspection of these latter seven spectra shows that they indeed appear to have resolutions well above 10,000. The $\mathcal R=25,000$ fits were retained as the adopted inclination angles for these seven stars (see Table~\ref{table:sample}); however, as we judge our minimum fit uncertainty to be $\pm5^\circ$, these $4^\circ-6^\circ$ inclination differences caused by the resolution of the model profiles were not significant.

\ha\ fits for all 92 sample stars can be seen in Appendix~A, with the stars ordered via their HD number as in Table~\ref{table:sample}. Shown for each star is the observed profile, and the four model profiles corresponding to the four hyper-parameter best fits.

We note that \citet{Zorec2016} provides much more accurate stellar parameters (masses, radii, and luminosities) than can be obtaining by the crude spectral type calibration used in the \iha\ estimates here. However, we adopted the simple spectral type calibration for the \ha\ profile fitting in order to keep the two methods completely independent and to fairly compare the intended use of both methods. In the gravitational darkening method, the stellar inclination comes from detailed spectral synthesis designed to extract the fundamental stellar parameters. For the \ha\ method, it's intended use is for stars for which a moderate resolution \ha\ line profile and stellar spectral type are all that is available.

\section{Comparison of H$\alpha$ Fitting and Gravitational Darkening}

\begin{figure}
\begin{center}
\includegraphics[width=0.5\textwidth]{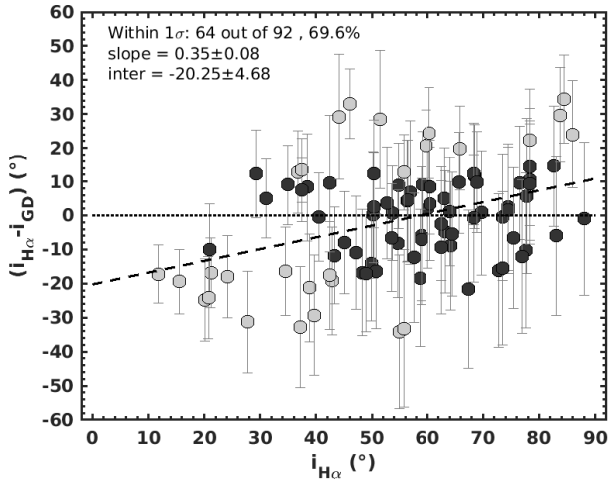}
\caption{The difference between the \ha\ and gravitational darkening inclinations versus the \ha\ inclinations. The errors bars are $1\sigma$. Of the 92 sample stars, 64 (or $70\%$) agree within the errors and are represented by the darker circles. The dashed line is a least-squares fit to the data and the slope and y-intercept, plus their bootstrap Monte Carlo errors, are given in the figure. \label{fig:bvz_diff}}
\end{center}
\end{figure}

Table~\ref{table:sample} lists the results of \ha\ line profile fitting ($i_{\rm H\alpha}\pm\,1\sigma$) and gravitational darkening ($i_{\rm GD}\pm\,1\sigma$) \citep{Zorec2016} for each of the 92 sample stars. Figure~\ref{fig:bvz_diff} plots the inclination angle difference, $(i_{\rm H\alpha}-i_{\rm GS})$ versus \iha\ with the error bars representing the combined uncertainty of the individual errors added in quadrature. As can be seen from the figure, 64 out of the 92 stars (70\%) are consistent with zero difference between the angles, suggesting that the $1\sigma$ error bars of the two methods are realistically determined. However, there is a trend in the differences in that for lower inclinations, the gravitational darkening inclinations tend to be smaller than the \ha\ ones, whereas at high inclinations, the trend is the opposite with the \ha\ inclinations tending to be larger than the gravitational darkening ones. A straight-line fit to differences $(i_{\rm H\alpha}-i_{\rm GS})$ versus \iha\ leads to a positive and statistically significant slope of $0.35\pm0.08$, with the uncertainty estimated from boot-strap Monte Carlo. This trend is discussed below.

\begin{figure}
\begin{center}
\includegraphics[width=0.5\textwidth]{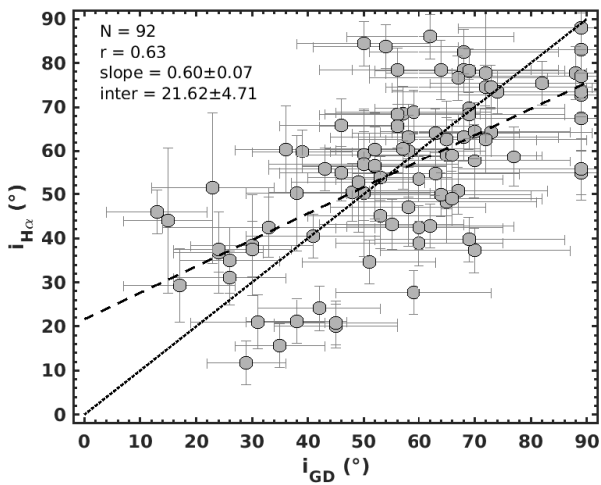}
\caption{A  comparison  of  the  sample  inclination  angles found  from  \ha\ profile-fitting  and  gravitational  darkening. The error bars are 1$\sigma$.  The dotted line is of unit slope, and the dashed line is a least-squares fit to the data.  The slope and y-intercept, with bootstrap Monte Carlo errors, are given in the figure.\label{fig:bvz}}
\end{center}
\end{figure}

A plot of \iha\ versus \igd\ is shown in Figure~\ref{fig:bvz}. The trend noted previously is also clearly visible in this figure, with the best-fit line to the data having a slope noticeably below one. 
The correlation coefficient between \iha\ and \igd\ is seen to be $r=+0.63$. A boot-strap Monte Carlo analysis gives $r=+0.63\pm0.05$. To assess the statistical significance of this result, we have performed $10^5$ simulations in which angles $i_1$ and $i_2$ were independently and randomly chosen from the $\sin i$ distribution for a sample of 92 stars. Over the $10^5$ simulations, the min and max correlation coefficients were $-0.412$ and $+0.435$, respectively, and the distribution of $r$ is well-fit by a Gaussian of mean zero and $\sigma=0.105$. Thus the observed correlation between \iha\ and \igd\ is approximately six standard deviations from the mean of 0 expected from zero correlation.

\begin{figure}
\includegraphics[width=0.5\textwidth]{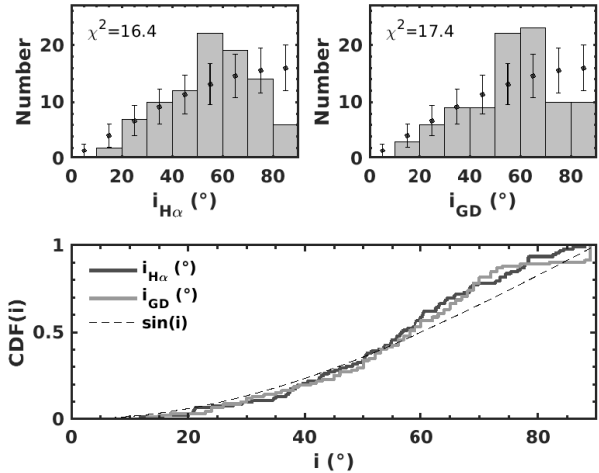}
\caption{The distribution of \iha\ (top left) and \igd\ (top right), in bins of $10^\circ$, compared to the number expected in each bin based on the random $\sin i$ distribution. The $\chi^2$ of each sample is indicated in each panel. The bottom panel compares the cumulative distribution (CFD) of both \iha\ and \igd\ with the CDF expected for the random $\sin i$ distribution, $(1-\cos i)$.
\label{fig:cdf}}
\end{figure}

Figure~\ref{fig:cdf} (top panels) shows the sample inclination distributions for both the \iha\ and \igd\ in bins of $10^\circ$ in width. Also shown is the number of stars expected in each of the 9 bins for the $\sin i$ distribution. The $\chi^2$ for the \ha\ fitting method is 16.4, expected to occur by chance only $3.8$\% of the time if the underlying distribution were the random $\sin i$ distribution. The $\chi^2$ for the gravitational method is slightly larger at 17.4, expected to occur by chance only $2.6$\% of the time.\footnote{The 92 sample stars represent less half of the full 233 star \citet{Zorec2016} sample. Using the same 9 inclination bins, the full sample has $\chi^2=34.2$ with a probability of $0.004$\% of occurring by chance under the random $\sin i$ distribution.} Thus, there is marginal statistical evidence that the observed sample inclination distribution is not random. Both distributions have a lack of high inclination objects with either $i > 80^\circ$ if one considers the \ha\ determinations or $i> 70^\circ$ using the gravitational darkening determinations.\footnote{The deficit of stars with $i\ge 70^\circ$ is even more pronounced in the full 233 star \citet{Zorec2016} sample.} This tendency of observed samples of Be stars to lack high inclination objects (Be shell stars) has been noticed and discussed before \citep[see][]{Rivinius2006a}. This may be a selection effect: a common way to select Be star candidates for spectroscopic confirmation is to look for B-type stars with an infrared excess due to free-free emission from the disk. However, for systems in which the star is viewed through the disk, i.e.\ shell systems, the excess may be negligible or absorption of the photospheric spectrum may even lead to a deficit \citep[see, for example,][]{SigutPatel2013}. Shell stars and the apparent lack of high inclination objects are discussed in somewhat more detail in the next section.

Finally, the bottom panel of Figure~\ref{fig:cdf} shows the inclination distributions produced by both methods as cumulative distributions (CDFs). A Kolmogorov-Smirnov (KS) test for the expected $(1-\cos i)$ CDF corresponding to the random $\sin i$ distribution leads to results consistent with the $\chi^2$ analysis given above. 

\section{Trend with the H$\alpha$ Shell Parameter}

An important class of Be stars are Be shell stars in which the \ha\ line centre absorption is very deep, well below the absorption expected from the central B star's photospheric profile. The interpretation of this narrow, central absorption is that the observer is looking through the disk to the central star and the absorption core corresponds to absorption of the photospheric spectrum by the disk \citep{Porter2003}. As Be star disks are very thin, shell stars must be seen at high inclination angles ($i\gtrsim 75^\circ)$ by the observer, a fact supported by the systematically higher $v\sin i$ for Be shell stars \citep{Porter1996}. This view is further supported by the observed correlations between H$\alpha$ emission line strength and continuum magnitude during Be disk building or dissipation phases \citep{Harmanec1983}. While most Be stars show a positive correlation, continuum brightening as the H$\alpha$ emission strength increases, shell stars typically show an inverse correlation, i.e.\ a dimming as the H$\alpha$ emission increases. This inverse correlation is expected if shell stars are viewed at high inclination as the building disk obscures the star \citep{Harmanec1983,Harmanec2000,SigutPatel2013}. 

\citet{Hanuschik1996} refined the observational definition of a Be shell star by introducing the {\it shell parameter,} defined as the average of the flux in the violet and blue emission peaks in \ha\ divided by the central flux. \citet{Hanuschik1996} carefully defined disk absorption in terms of weak, Fe\,{\sc ii} lines in the spectrum and found that such absorption occurs when the H$\alpha$ shell parameter first exceeds 1.5. We adopt this definition here.

\begin{figure}
\includegraphics[width=0.5\textwidth]{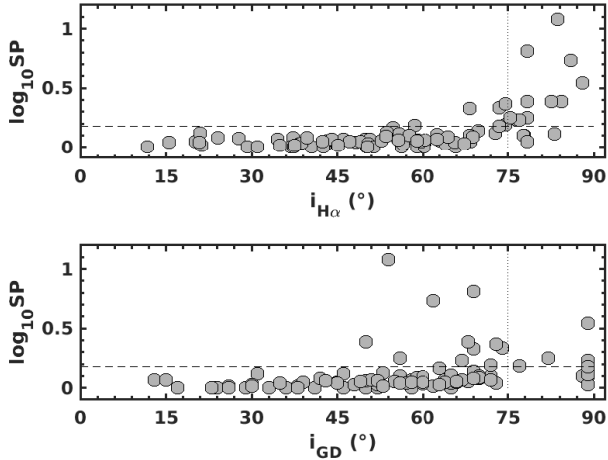}
\caption{The \ha\ shell parameter (Table~1) versus \iha\ (top panel) and \igd\ (bottom panel). The horizontal dashed line in each panel corresponds to the critical shell parameter of 1.5 \protect\citep{Hanuschik1996}. The vertical dashed line at $75^\circ$ is to guide the eye. \label{fig:shellp}}
\end{figure}

As the shell parameter is easily measured for each star in our sample (see Table~\ref{table:sample}), we have the prediction that stars with measured shell parameters above 1.5 should be seen at high inclination angles, $i\gtrsim 75^\circ$. Figure~\ref{fig:shellp} tests this prediction using both the \ha\ and gravitational darkening inclination angles. Overall the \ha\ inclination angles satisfy this constraint with all shell stars of Table~\ref{table:sample} are seen to occur at high inclination angles (note that the inclination error bars are not shown in this figure for clarity). The gravitational darkening inclination angles, however, do not show this trend, with several stars with large shell parameters predicted to be viewed at intermediate inclination angles. 

Revisiting Figure~\ref{fig:cdf}, some of the high-inclination stars for the gravitational darkening inclinations may be mis-classified to more intermediate inclinations, leading to the structure seen in this histogram. However, there is also a deficit in the highest inclination bin for the H$\alpha$ inclinations. If the expected number of shell stars were present in the sample, the H$\alpha$ fitting method would almost certainly have correctly identified them; hence, we conclude that sample itself has a selection effect in which high inclination objects are underrepresented, for the reason noted previously. We feel that the structure in the gravitational darkening inclination distribution results from both of these effects: there is a real deficit of high-inclination stars, combined with some mis-classifications to lower inclinations.\footnote{This extends to the full 233 star \citet{Zorec2016} sample as well, where the deficit of high inclination objects is even more extreme.}

However, there is a potential caveat to this conclusion. The
occurrence of shell absorption in the model profiles is governed by the
disk thickness set by the scale height of Eq.~(\ref{eq:H}). If this scale
height were an overestimate, then the occurrence of shell absorption in
the models would be artificially shifted to lower inclinations, perhaps
contributing to the observed deficit of high inclination objects. In
Figure~\ref{fig:shellp}, there is not a large number of stars with a
\ha\ shell parameter close to the critical value of 1.5 but assigned
to $i_{\rm H\alpha} < 75^\circ$; however, reclassification of only
$\approx 5$ stars to $i\gtrsim 80^\circ$ (doubling the number in the
final $80-90^\circ$ bin) would make the observed numbers more consistent
with the $\sin i$ distribution. This idea, that the disks are slightly
too thick, is supported by the calculations of \citet{Sigut2009} where
the disk scale height is not determined by Eq.~(\ref{eq:H}) but by
integrating the equation of hydrostatic equilibrium vertically at each
disk radius to ensure consistency with the temperature distribution. As
Be star disks often possess a cool, equatorial zone near the star
where all rays back to the star have significant optical depth
\citep{Carciofi2006,Sigut2007}, the inner disk scale heights are often
less than predicted by Eq.~(\ref{eq:H}). However, it is not clear if
this effect is large enough to explain the deficit of high inclination
objects. We will leave a re-analysis of this sample with a grid of
disk models in consistent, vertical hydrostatic equilibrium to future
work.\footnote{Disk models in consistent, vertical hydrostatic
equilibrium typically take significantly longer to compute than those based
on Eq.~(\ref{eq:H}) as one must iterate the temperature and density
distributions to achieve convergence \citep[see][for details]{Sigut2009}}.

\section{The Equatorial Velocity Distribution}

Another consistency test is possible, this time primarily at lower inclination angles. The projected rotation velocity, $v\sin i$, is directly measurable from the broadening of photospheric lines. If an independent estimate of the inclination angle of the rotation axis is available, we can extract the true equatorial rotation velocity directly from $v\sin i$. As Be stars are well known to be rapid rotators, we expect a strong correlation between $v\sin i$ and $i$ in the sense that low $v\sin i$ should be associated with low $i$ so that the star's equatorial rotation speed is consistent with it's nature as a Be star. In addition, the upper limit to $v$ is the critical velocity, Eq.~(\ref{eq:vcrit}), and we should not find stars that appear to be rotating above this limit. To this end, we have taken the measured $v\sin i$ of \citet{Zorec2016} and combined them with both the H$\alpha$ and GD inclination estimates to extract the equatorial velocity distribution. 

It is well-known that $v\sin i$ determinations for Be stars must account for the effects of gravitational darkening, otherwise the  $v\sin i$ may be underestimated due to the reduced contribution of the fastest rotating, equatorial regions \citep{Townsend2004a}; the reduction can be significant if a particularly temperature sensitive line like He\,{\sc i} $\lambda\,4471\,$\AA\ is used to determine $v\sin i$. In the present work, we simply adopt the $v\sin i$ estimates of \citet{Zorec2016} which are (obviously) derived including gravitational darkening in the analysis. These $v\sin i\pm 1 \sigma$ values are given in Table~\ref{table:sample}.

\begin{figure}[t]
\includegraphics[width=0.5\textwidth]{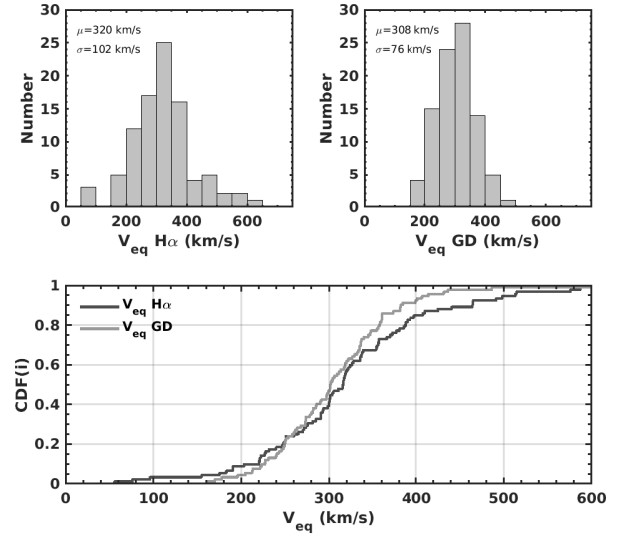}
\caption{The tow top panels show the histograms of stellar equatorial rotational speed derived from the $v\sin i$ values of \citet{Zorec2016} using \iha\ (left panel) and \igd\ (right panel). The bottom panel shows the corresponding cumulative distributions.\label{fig:veq}}
\end{figure}

The top two panels of Figure~\ref{fig:veq} show the sample star equatorial rotation speeds using \iha\ (left) and \igd\ (right) to recover $v$ from $v\sin i$. Both rotation speed distributions are consistent with a Gaussian distribution as revealed by a KS test. The means and widths of both distributions are quite similar, $v\pm 1\sigma$ equal to $320\pm 102\,\rm km\,s^{-1}$ for \iha\ and $308\pm 76\,\rm km\,s^{-1}$ for \igd. The bottom panel of Figure~\ref{fig:veq} shows the cumulative distributions of the equatorial rotation speeds and a KS test accepts the null hypothesis that the two distributions are the same. Nevertheless, there are differences: the \iha\ speed distribution shows a small number of stars with recovered equatorial speeds below $200\,\rm km\,s^{-1}$, troubling for a population which are all expected to be rapid rotators. In addition, the \iha\ equatorial speed distribution has a small number of stars in an extended high-velocity tail, not present in the \igd\ sample, with speeds that seem to exceed the critical velocity. However, both of these low-speed and high-speed stars need to viewed within the context of the errors in the derived equatorial speeds, as we shall now discuss. 

\begin{figure}[t]
\includegraphics[width=0.5\textwidth]{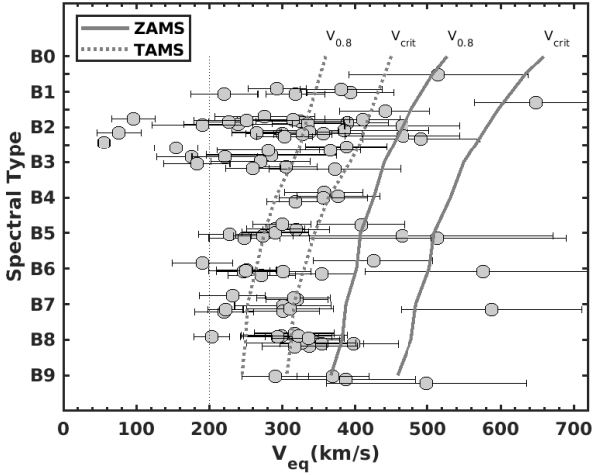}
\caption{Equatorial rotation speed derived using the $v\sin i$ values of \citet{Zorec2016} and \iha\ for the sample stars as a function of spectral type. The critical rotational speed and 80\% of the critical speed are shown for both the zero-age (ZAMS) and terminal-age (TAMS) main sequence. \label{fig:spt_veq}}
\end{figure}

To place the ``low speed" ($v\lesssim 200\,{\rm km\,s^{-1}}$) and the ``high speed" ($v\gtrsim 500\,{\rm km\,s^{-1}}$) rotators based on \iha\ in context, Figure~\ref{fig:spt_veq} shows the stellar equatorial rotation speed of the sample stars, with $1\sigma$ errors based on the combined uncertainties in $v\sin i$ and \iha, as a function of the star's spectral type. To guide the eye, the critical rotational velocity, Eq.~(\ref{eq:vcrit}), and 80\% of the critical velocity, for both the ZAMS and TAMS are shown as a function of spectral type. The masses and radii as a function of spectra type for the ZAMS and TAMS used to compute the critical velocities were taken from the models of \citet{Georgy2012}. 

First, we note that all of the high speed stars have very large errors and none are inconsistent with critical rotation. For the low-speed stars, only three stand out as being incompatible with $200\,\rm km\,s^{-1}$, the low-speed edge of the \igd\ equatorial speed distribution see in Figure~\ref{fig:veq}: HD~44996 ($56\pm 7\,\rm km\,s^{-1}$), HD~58343 ($96\pm 30\,\rm km\,s^{-1}$), and HD~167128 ($76\pm 30\,\rm km\,s^{-1}$). These ``slow rotators" do not occur with \igd\ determined inclinations as \igd\ is significantly smaller for each star. As these very low $v$ values seem implausible, we feel that the error lies in the \iha\ values at low inclinations, something noted before.

We note that the average rotation speed found for $v$ previously using \iha, $320\,\rm km\,s^{-1}$ seems representative of all spectral types. Breaking the same into early-type (B3 or earlier) and late-type (B4 and later) gives very similar means, $310\,\rm km\,s^{-1}$ and $329\,\rm km\,s^{-1}$, respectively. However, the sample average is reduced somewhat if a $1/\sigma$ weighted-mean is taken; in this case the sample weighted mean is $282\,\rm km\,s^{-1}$.

Finally, we have not attempted to determine the $v_{\rm frac}\equiv v/v_{\rm crit}$ distribution for the sample; a meaningful investigation of this distribution requires more accurate masses and radii for the sample stars than an average spectral type calibration can provide; we defer to the results of \citet{Zorec2016} in this case.

\section{Disk Density Parameters for the Sample Stars}
\label{sec:otherparam}

\begin{figure}[t]
\vspace{0.5cm}
\includegraphics[width=0.45\textwidth]{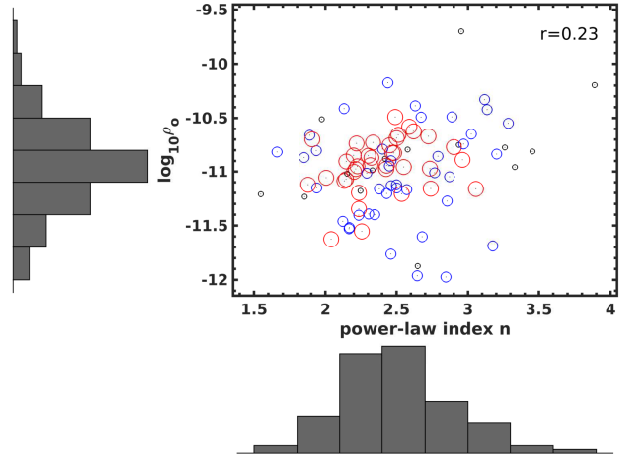}
\caption{Disk density parameters $(n,\rho_0,R_d)$ of the best-fit \ha\ profile for the 92 sample stars. Each disk is shown as a point in the $n-\log_{10}\rho_0$ plane. The symbol size and colour represents $R_d$, with $R_d\le15$ (smallest circles, black), $15<R_d<35$ (intermediate circles, blue) and $R_d>35$ (largest circles, red). The disk radius is in units of the stellar radius. Relative histograms for the $n$ and $log_{10}\rho_0$ distributions are also shown. \label{fig:rhon}}
\end{figure}

A natural bi-product of the \iha\ fitting method is estimates for the disk density parameters: $\rho_0$, the density of the inner edge of the disk in the equatorial plane, $n$, the index in the power-law drop-off in density as a function of distance from the rotation axis, and $R_D$, the outer radius of the disk (measured in stellar radii). The distribution of these parameters are shown in Figure~\ref{fig:rhon}. The range of values found is consistent with other studies, with $10^{-12} \lesssim\rho_0({\rm g\,cm^{-3}})\lesssim 10^{-10}$ and $n$ in the range of 1.5 to 4 \citep{Silaj2014,Vieira2017}. There is a slight trend in the data in that the $\log\rho_0$ and $n$ values are correlated with $r=+0.23$. Part of this correlation may be due to the selection effect associated with undetectable disks: one would expect any disks in the lower right of the figure, low $\rho_0$ and high $n$, to produce little emission and be undetectable in \ha, and one would expect this limit to include a correlation between $\rho_0$ and $n$ in the sense that the detection limit moves to larger $\rho_0$ for larger $n$, {\it i.e.}\ a positive correlation. 

\section{Discussion}

We have demonstrated that stellar inclination angles determined for Be stars from either the modelling of \ha\ emission arising from the {\it circumstellar\/} disk (\iha) or the modelling of the spectroscopic signature of {\it photospheric\/} gravitational darkening (\igd) agree with one another to within the estimated errors. For a sample of 92 stars, (\iha - \igd) is found to be consistent with zero within $\pm 1\sigma$ for 70\% of the sample stars. These two methods are completely independent, up to and including the observational data used. The gravitational darkening method uses optical, photospheric spectra, excluding lines exhibiting circumstellar emission, with the inclination angle found as part of an overall analysis to determine fundamental stellar parameters. The \ha\ method uses a spectrum of the \ha\ emission line, with the inclination angle found by matching the observed line profile with a \ha\ template library specific to the spectral type of the star.

\begin{figure}[t]
\includegraphics[width=0.5\textwidth]{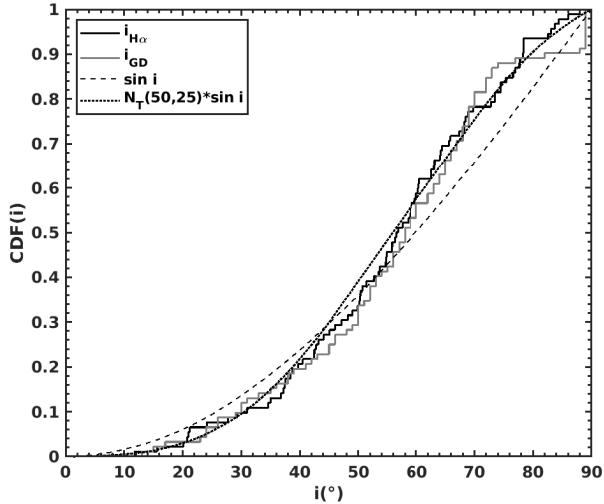}
\caption{The cumulative distributions of \iha\ (black solid line) and \igd\ (grey solid line) for the 92 sample Be stars. Shown also is the cumulative distribution of the random $\sin i$ distribution (dashed line) and the cumulative distribution for the best-fit truncated Gaussian of Eq.~(\ref{eq:nt}) (dotted line), $\mu=50^\circ$ and $\sigma=25^\circ$. This best-fit was found following the procedure outlined in \citet{Sigut2020}. \label{fig:ntfit}}
\end{figure}

To use either of these methods to search for correlated stellar spins in young, open clusters, particularly for cases of moderate alignment, a careful understanding of the inherent biases of the methods is required. This work presents evidence that the gravitational darkening method likely underestimates the inclination for $i>70^\circ$, perhaps due to the neglect of the disk on the optical spectrum used in the analysis. For \ha\ method, there is evidence that the inclination is over-estimated for inclinations $i<25^\circ$, perhaps due to the neglect of the intrinsic line broadening produced by inelastic electron scattering \citep{Poeckert1979}. Work is currently underway to recompute the \ha\ line libraries including this process.

Another potential issue is bias in observed Be star samples. The current work builds on the conclusion that field samples of Be stars are deficit in high-inclination objects (Be shell stars), likely due to the photometric identification of Be star candidates for spectroscopic confirmation. The effect of a bias such as this on searches for moderate spin correlation is shown in Figure~\ref{fig:ntfit}. Here the cumulative distributions of \iha\ and \igd\ for the 92 sample stars of this work are fit with the cumulative distribution predicted by a probability function of the form 
\begin{equation}
\label{eq:nt}
p(i)\,di\equiv {\cal N} N_{\rm T}(\mu,\sigma)\,\sin i\,di \;.
\end{equation}
Here $N_{\rm T}(\mu,\sigma)$ is a truncated Gaussian\footnote{The truncated Gaussian was implemented with the \texttt{Matlab} R2019a \texttt{truncate} function which sets the Gaussian PDF to zero for $\mu<0^\circ$ and $\mu>90^\circ$ and re-normalizes it to unit area.} of mean $\mu$ and standard deviation $\sigma$, and ${\cal N}$ is a constant to ensure unit normalization of $p(i)$ over the (truncation) interval $0^\circ\le\mu\le\,90^\circ$. While ad-hoc, this equation is a simple way to parameterize correlated spins \citep[see][]{Sigut2020}.  Fitting to the observed inclination distributions of Figure~\ref{fig:ntfit}, one finds that $\mu=50^\circ$ and $\sigma=25^\circ$ provides the best fit to the data, suggesting weak but detectable spin alignment. Of course, this is likely spurious as the observational sample is biased against high inclination objects; however, it is a good illustration of the strict requirement for unbiased samples. Fortunately, spectral synthesis for the Be stars is sufficiently well developed that codes such as \texttt{Bedisk} and \texttt{Beray} can be used to theoretically estimate the level of bias and, perhaps, suggest methods to avoid the selection bias in the first place. Work is also underway to address this issue.   

\section*{Acknowledgements}
We would like to thank the referee, Dr.\ Juan Zorec, for helpful comments. T.\ A.\ A.\ S.\ acknowledges support from the Natural Sciences and Engineering Council of Canada through a Discovery Grant.


\appendix

\begin{figure}
\includegraphics*[width=\textwidth]{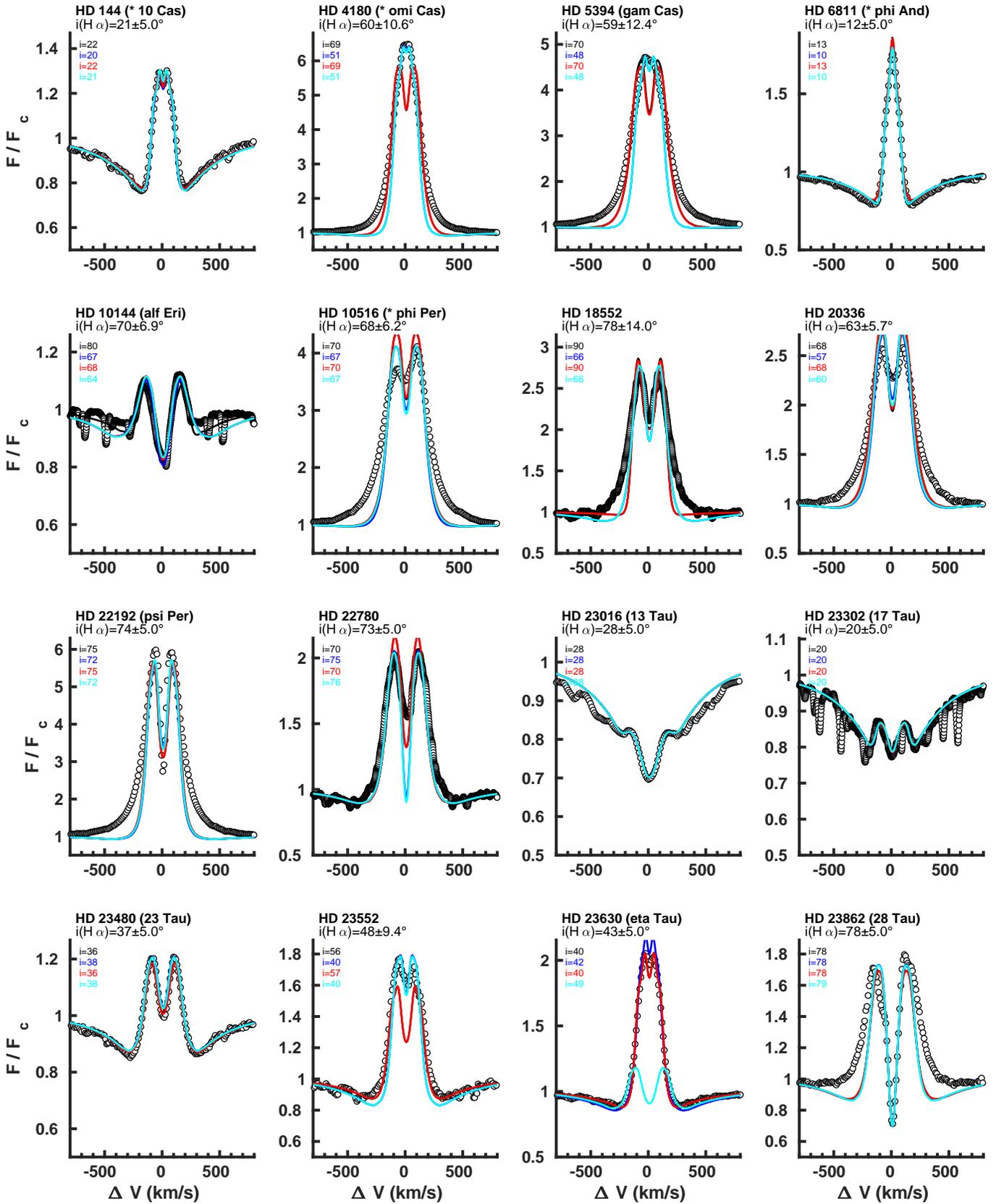}
\caption{\protect\ha\ profile fits to the sample stars. \label{fig:fits1}}
\end{figure}

\begin{figure}
\includegraphics*[width=\textwidth]{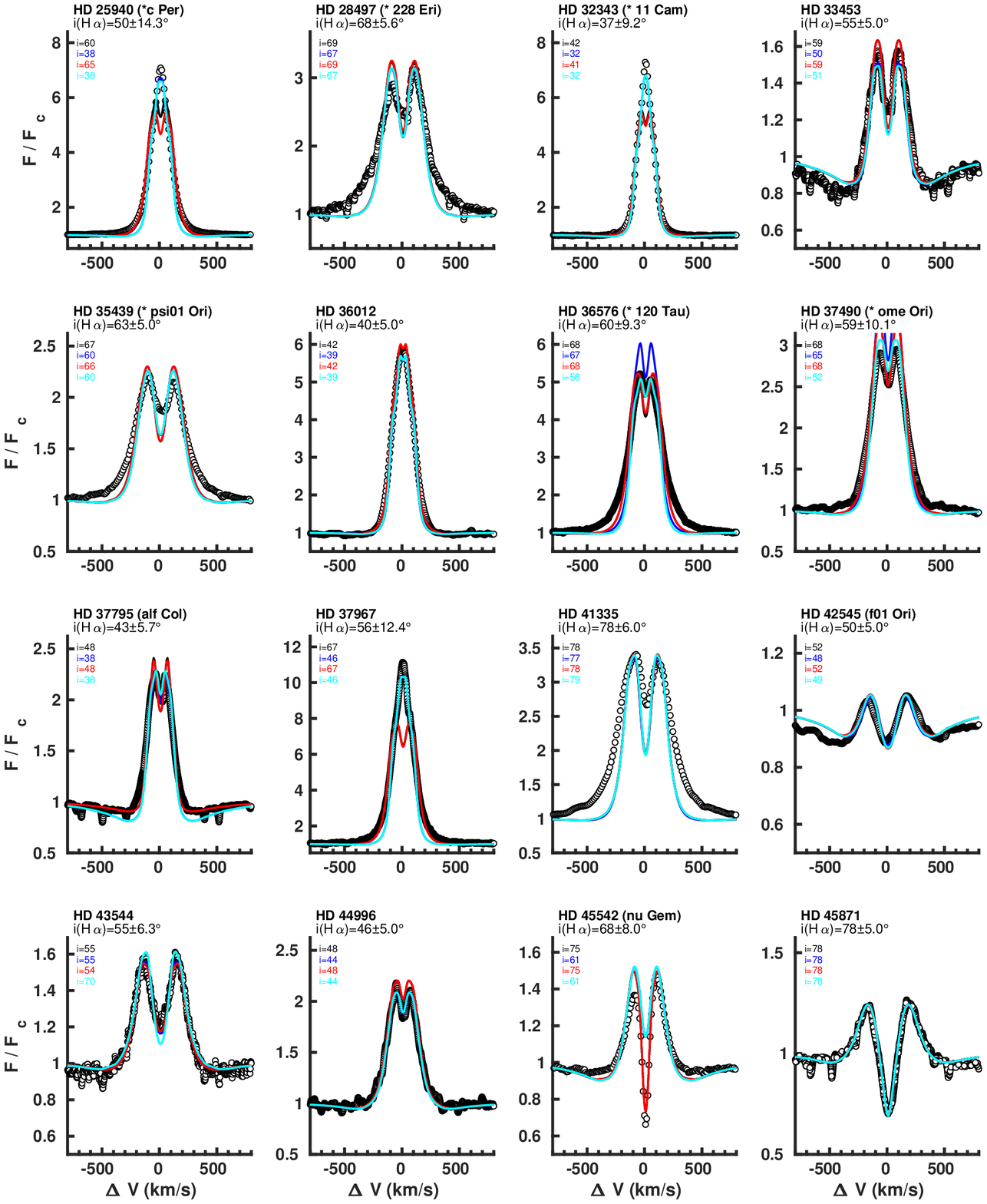}
\caption{\protect\ha\ profile fits to the sample stars continued. \label{fig:fits2}}
\end{figure}

\begin{figure}
\includegraphics*[width=\textwidth]{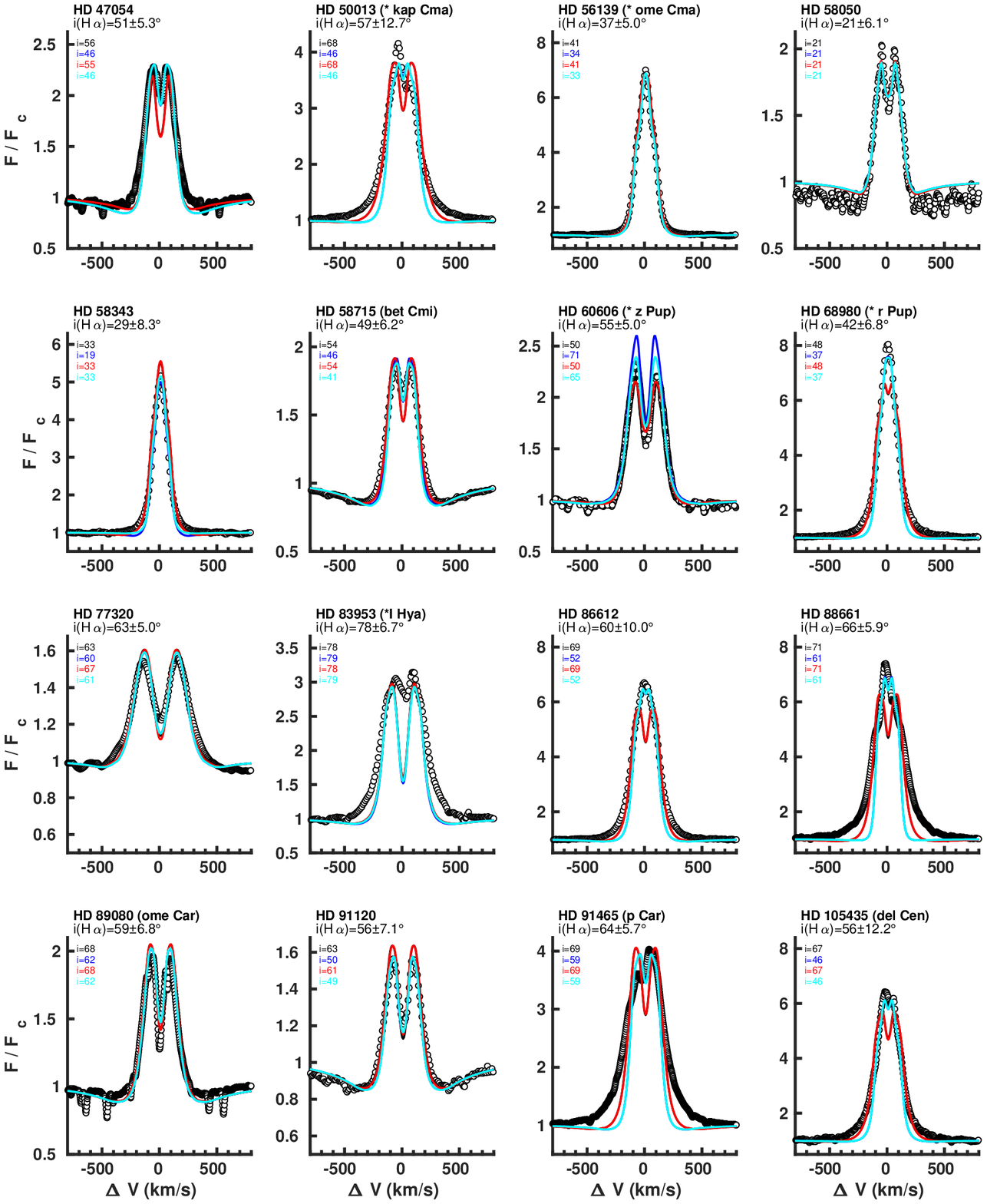}
\caption{\protect\ha\ profile fits to the sample stars continued. \label{fig:fits3}}
\end{figure}

\begin{figure}
\includegraphics*[width=\textwidth]{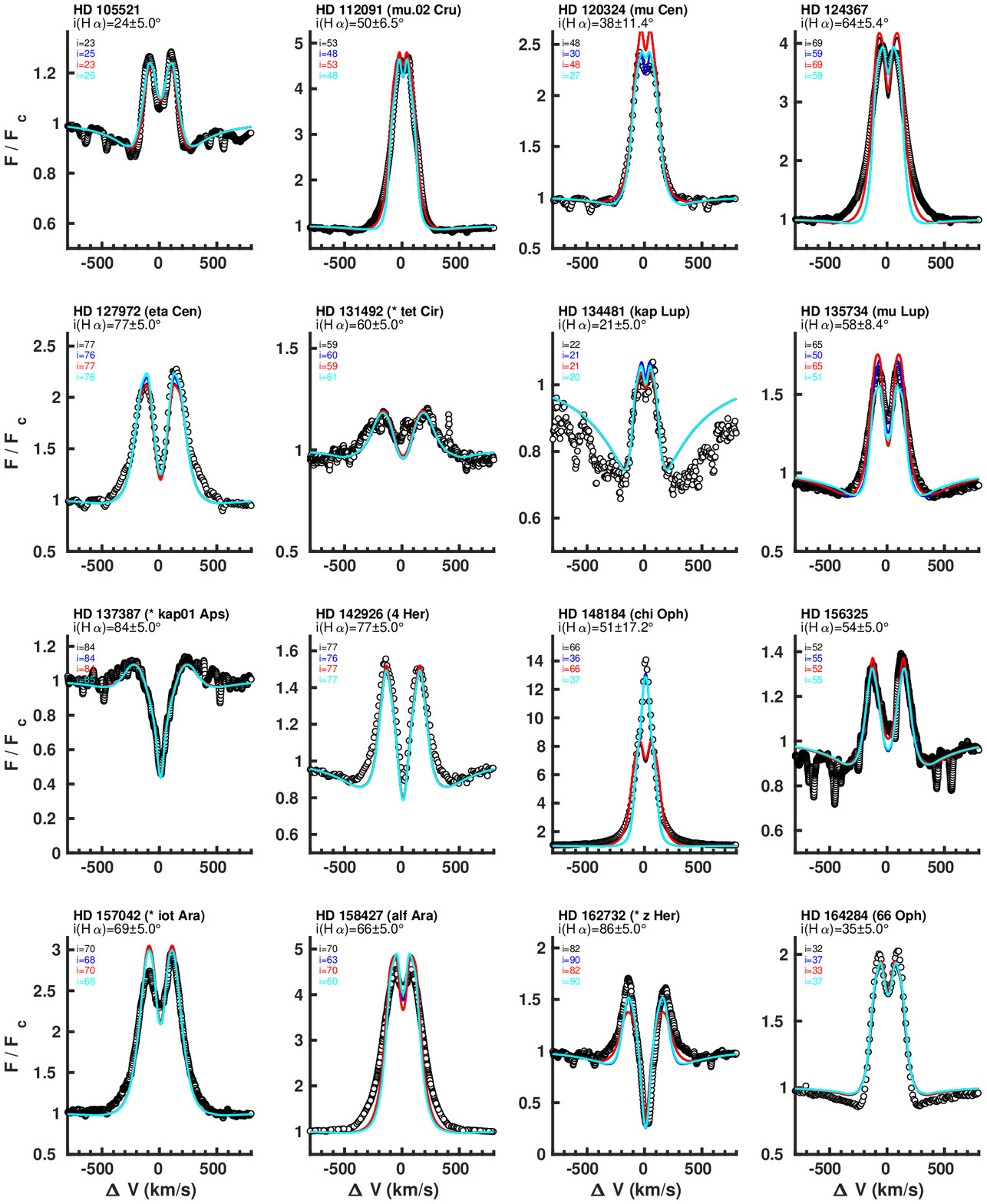}
\caption{\protect\ha\ profile fits to the sample stars continued. \label{fig:fits4}}
\end{figure}

\begin{figure}
\includegraphics*[width=\textwidth]{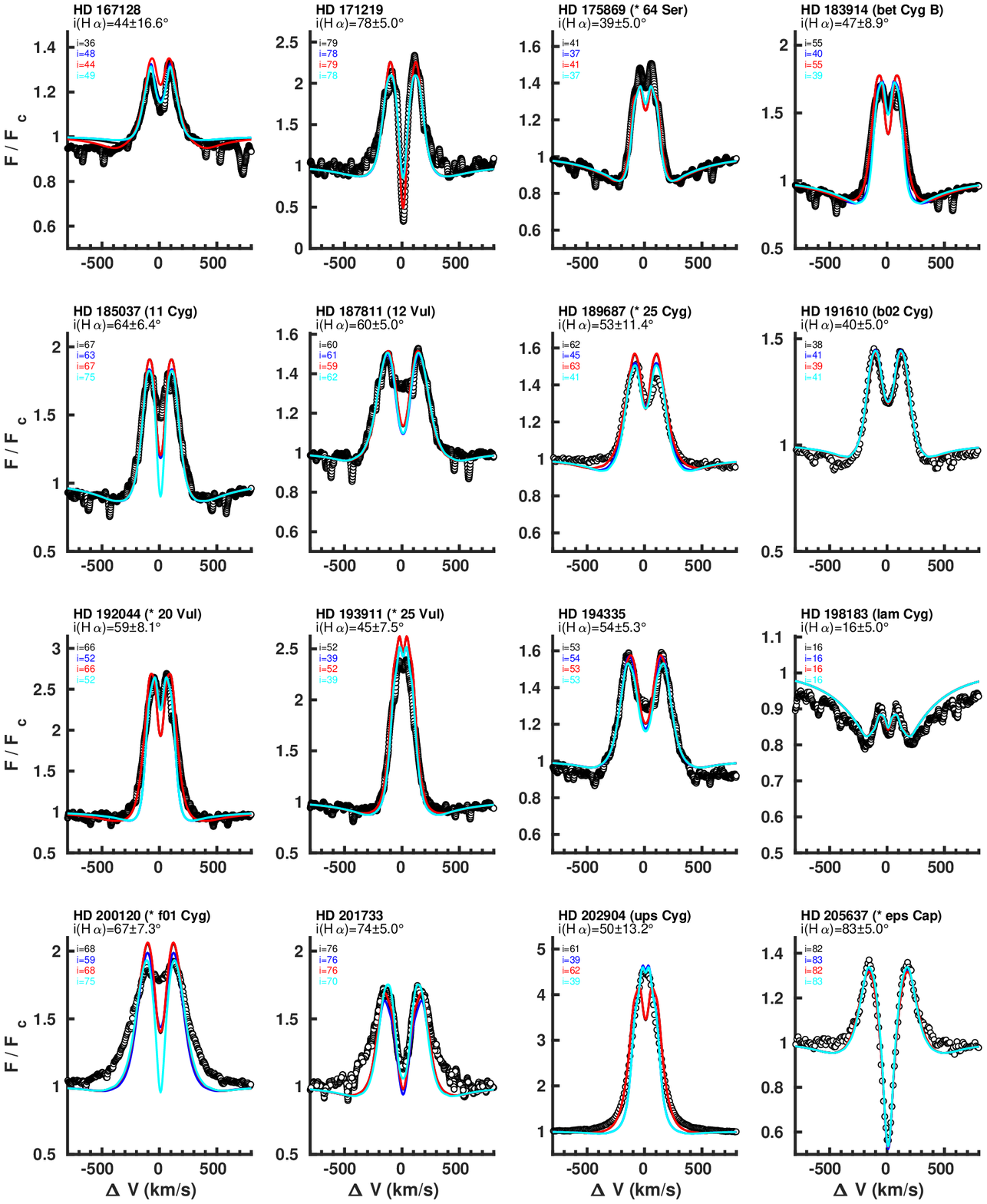}
\caption{\protect\ha\ profile fits to the sample stars continued. \label{fig:fits5}}
\end{figure}

\begin{figure}
\includegraphics*[width=\textwidth]{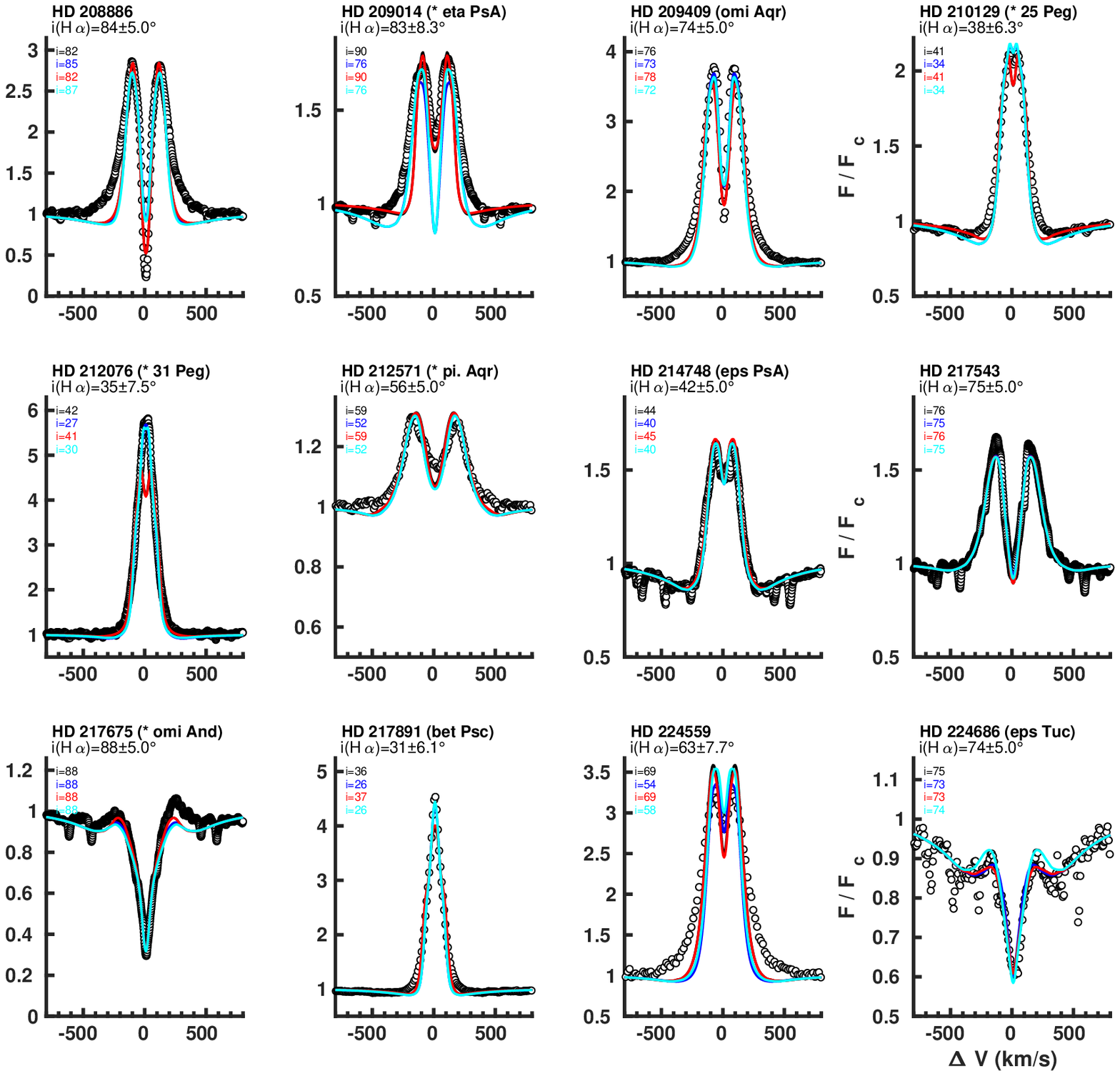}
\caption{\protect\ha\ profile fits to the sample stars continued. \label{fig:fits6}}
\end{figure}

Figures~\ref{fig:fits1} through \ref{fig:fits6} show the H$\alpha$ model fits to the sample stars. The stars are arranged in order of HD~number as in Table~\ref{table:sample}. Shown for each star are the observations (open circles) and four \ha\ profile fits corresponding to the four choices for the fit ``hyper-parameters:" the wavelength range used for the fit and the form of figure-of-merit used to measure the quality of the fit: uniform weighting, $\Delta\lambda=15\,$\AA\ (black line); core weighting, $\Delta\lambda=15\,$\AA\ (blue line); uniform weighting, $\Delta\lambda=12\,$\AA\ (red line); core weighting, $\Delta\lambda=15\,$\AA\ (cyan line). The inclination angle found from the fitting procedure in each case is shown in the upper left following the same colour scheme as the line. The adopted \iha\ and its $1\sigma$ error are also given in each panel. 

\newpage

\bibliography{citas.bib}

\end{document}